\documentclass[12pt]{article}
\usepackage[utf8]{inputenc}
\usepackage[english]{babel}
\usepackage{cmap}
\usepackage{authblk}
\usepackage{amssymb}
\usepackage{amsmath,mathtools}
\usepackage{amsfonts}
\usepackage{cite}
\usepackage{floatflt}
\usepackage[pdftex]{graphicx}
 
\usepackage{eeAA_definitions}
\newcommand{\SANC}{\texttt{SANC}{ }}

\newcommand{\MCSANCee}{\texttt{MCSANCee}{ }}
\newcommand{\ReneSANCe}{\texttt{ReneSANCe}{ }}
\newcommand{\WHIZARD}{\texttt{WHIZARD}{ }}
\newcommand{\CalcHEP}{\texttt{CalcHEP}{ }}
\newcommand{\BabaYaga}{\texttt{BabaYaga}{ }}

\usepackage{slashed}
\usepackage{Spinor} 
\newcommand{\lGAgrade}{\hstretch{0.6}{\prec} }
\newcommand{\rGAgrade}{\hstretch{0.6}{\succ} }
\newcommand{\grade}[2]{\stretchrel*{\lGAgrade }{#2|}#2\stretchrel*{\rGAgrade}{#2|}_{#1}}
 \DeclareMathOperator{\Tr}{Tr}  
 \DeclareMathOperator{\Trbar}{\overline{Tr}} 

\begin{document}
\title{\LARGE 
 One-loop radiative corrections to photon-pair production
 in polarized positron-electron annihilation 
}
\author[1]{S.\,Bondarenko}
\author[2,3]{Ya.\,Dydyshka}
\author[2]{L.\,Kalinovskaya}
\author[2]{A.\,Kampf}
\author[2]{L.\,Rumyantsev}
\author[2]{R.\,Sadykov}
\author[2,3]{V.\,Yermolchyk}

\affil[1]{\small Bogoliubov Laboratory of Theoretical Physics, JINR, 
                 141980 Dubna, Moscow region, Russia}
\affil[2]{\small Dzhelepov Laboratory of Nuclear Problems, JINR,  
                 141980 Dubna, Moscow region, Russia}
\affil[3]{\small Institute for Nuclear Problems, Belarusian State University, Minsk, 220006  Belarus}

\maketitle

\begin{abstract}
A theoretical description of photon-pair production in polarized positron-electron annihilation
is presented.
Complete one-loop electroweak radiative corrections are calculated 
taking into account the exact dependence on the electron mass.
Analytical results are derived with the help of the \SANC~system.
The relevant contributions to the
cross section are calculated analytically using the helicity amplitude approach.
The cases of unpolarized and longitudinally polarized fermions in the initial state are investigated.
Calculations are realized in the Monte Carlo integrator \MCSANCee and generator \ReneSANCe
which allow one the implementation of any experimental cuts used in the analysis of $e^+e^-$ annihilation data of both low and high energies.

\end{abstract}

\section{Introduction}
The comprehensive long-term program 
 of next generation 
\(e^{+}e^{-}\) colliders 
proposes a large potential improvement in ultra-precise 
measurements of electroweak (EW) parameters and 
the creation of modern tools for adequate luminosity estimates.
Future 
\(e^{+}e^{-}\) 
colliders 
such as
\texttt{FCC-ee}\cite{Abada:2019lih}, \texttt{ILC}\cite{Behnke:2013xla}, \texttt{CEPC}\cite{CEPCStudyGroup:2018rmc,CEPCStudyGroup:2018ghi} and \texttt{CLIC}\cite{Aicheler:2012bya} will have a total luminosity 2-3 orders of magnitude larger than the LEP total luminosity  
and the possibility of using polarizing beams that could provide an additional probe of the accuracy test of the Standard model as well as in the search for new physics.
The determination of the luminosity at lepton colliders is a necessary task, since the normalization of measured cross sections is an observable quantity of immediate phenomenological interest. 
Permille precision
of the integrated luminosity measurement at future colliders  seems 
feasible in terms 
of existing technologies
\cite{Smiljanic:2021bbb}.

In practice, this problem is solved by choosing three specific reference processes which
generate large statistics, are as free as possible from systematic ambiguities, and are predicted by theory with suitable accuracy:
small and large angle Bhabha scattering, 
 lepton-pair production in $e^{+} e^{-}$ collisions
and large angle $e^{+} e^{-}$ annihilation to photon pair.

The main result of this work is the independent calculation of
the complete one-loop EW radiative corrections (RCs) 
taking into account the exact dependence 
of the $e^{+} e^{-}$ annihilation to  photon pair
on the electron mass
\bqa
\label{cppp}
e^{+}(p_1,\chi_1)\ +\ e^{-}(p_2,\chi_2)\ \rightarrow 
 \gamma(p_3,\chi_3)\ +\ \gamma(p_4,\chi_4)\ \ (+ \gamma (p_5, \chi_5)),
\eqa 
and arbitrary longitudinal polarization of initial particles. 
Here $p_{i}$ are the 4-momenta and $\chi_i$ are the helicities of the corresponding particles.

The  
photon-pair production  plays a central role in the determination of the  luminosity for the following reasons:
events  have two collinear photons at large angles providing a clean signature;
the theoretical accuracy for the Bhabha process and the s-channel is 
limited by an uncertainty in the hadronic contribution $\Delta\alpha_5^{hadr}(s)$ to the vacuum polarization $\Pi_{\gamma\gamma}$,
but in the case of the process under consideration the hadronic contribution to the vacuum polarization enters only at the two-loop level
and the theoretical accuracy of $\Delta\alpha_5^{hadr}(s)$
is approximately of an order of
$10^{-6}$ \cite{CarloniCalame:2019dom}.

Process (\ref{cppp}) was first investigated
in the classical papers  
\cite{Brown:1952eu},
\cite{Harris:1957zza},
and later in
\cite{Berends:1973tm}.
Now it has already been studied at the one-loop level
in connection with experiments at future $e^+e^-$ colliders.
For the first time this was done in \cite{Berends:1973tm}.
Modern Monte-Carlo tools for the process ~(\ref{cppp}) are MC generators
{\tt MCGPJ} 
\cite{Eidelman:2010fu} 
and 
{\tt BabaYaga$@$NLO}~ \cite{CarloniCalame:2011aa,Calame:2014ucm,CarloniCalame:2017ioy,
CarloniCalame:2019dom}.
The recent 
version of the MC 
{\tt BabaYaga$@$NLO}~
contains one-loop calculations and also provides an enhancement of
leading logarithmic (LL) QED contributions due to multiphoton emission and the impact of photonic and fermion-loop corrections at next-to-next-to-leading order
\cite{CarloniCalame:2019dom}.

In the present paper, the calculations in the framework of {\tt SANC} are carried out  at the one-loop level within the OMS (on-mass-shell) renormalization
scheme in $R_\xi$ 
and in the unitary gauge as a cross-check.
Loop integrals are expressed in terms of the standard scalar
Passarino-Veltman functions \cite{Passarino:1978jh}.
To parameterize the ultraviolet divergences, dimensional regularization was  used.
Numerical results were obtained by the MC generator \ReneSANCe \cite{Sadykov:2020any} and integrator ~{\tt MCSANCee}.
To date, theoretical uncertainties  by {\tt SANC} have been investigated for the complete one-loop and leading higher-order EW corrections and with
taking into account polarization in the initial and final states 
for the following processes: 
        Bhabha scattering 
        \cite{Bardin:2017mdd}, 
 $e^{+} e^{-}$ annihilation to $Z H$ and $Z \gamma$
 \cite{Bondarenko:2018sgg}, 
s-channel 
\cite{Bondarenko:2020hhn},
        M\o{}ller scattering
        \cite{Bondarenko:2022kxq}, 
        and polarized $\mu e$ scattering 
        \cite{Arbuzov:2021oxs}.

This article consists of four Sections.
We describe the methodology of calculations of polarized cross sections at the complete one-loop EW level in the massive basis
within the helicity approach in Section~\ref{Sect_Css}.
The evaluation of theoretical uncertainties
for unpolarized~\texttt{FCCee},~\texttt{CEPC}  and 
polarized  \texttt{ILC}, \texttt{CLIC}
future experiments
and the results of a comprehensive comparison
of the independent MC codes for cross-checking are presented in  Section~\ref{Sect_Num}. 
~Summary is drawn in Section~\ref{Sect_Concl}.

\section{Differential cross section~\label{Sect_Css}}

The cross section of the process of the longitudinally polarized positron $e^+$
and electron $e^-$ beams with the polarization degrees $P_e^+$ and $P_e^-$,
respectively, can be written as follows:
\begin{equation}
\sigma{(P_{e^+},P_{e^-})} = \frac{1}{4}\sum_{\chi_1,\chi_2}(1+\chi_1P_{e^+})(1+\chi_2P_{e^-})\sigma_{\chi_1\chi_2},
\label{eq1}
\end{equation}
where $\chi_{1(2)} = -1(+1)$ correspond to the particle $i$ with the left (right) helicity.

The complete one-loop cross section of the process can be split into four parts
\begin{eqnarray}
  \sigma^{\text{one-loop}} =  \sigma^{\mathrm{Born}}
                       + \sigma^{\mathrm{virt}}(\lambda)
                       + \sigma^{\mathrm{soft}}({\lambda},{\omega})
                       + \sigma^{\mathrm{hard}}({\omega}).
\end{eqnarray}
Here $\sigma^{\mathrm{Born}}$ is the Born cross section,
$\sigma^{\mathrm{virt}}$ is the contribution of virtual (loop) corrections,
$\sigma^{\mathrm{soft(hard)}}$ is the soft (hard) photon emission contribution
(the hard photon energy {$E_{\gamma} > \omega$}).
The auxiliary parameters $\lambda$ ("photon mass")
and $\omega$ are canceled after summation.
Note that in calculations of one-loop RCs we can separate QED and pure 
weak interaction effects.

We apply the HA approach to all four components
of the one-loop cross sections.

The virtual (Born) cross section of the $e^+ + e^- \to \gamma + \gamma$ process
has the following form:
\bqa
\frac{d\sigma^{\mathrm{virt(Born)}}_{ \chi_1 \chi_2}}{d\cos{\vartheta}}
= \pi\alpha^2\frac{1}{4\beta_e}|\mathcal{H}^{\mathrm{virt(Born)}}_{\chi_1 \chi_2}|^2,
\eqa
where $\beta_e=\sqrt{1-4m_e^2/s}$, $\vartheta$ is the angle between the $p_1$ and $p_3$ and 
\bqa
|\mathcal{H}^{\rm virt(Born)}_{ \chi_1 \chi_2}|^2 =
\sum_{\chi_3,\chi_4} |\mathcal{H}^{\rm virt(Born)}_{\chi_1 \chi_2\chi_3\chi_4}|^2.
\eqa

The cross section for the hard photon bremsstrahlung 
is given by the expression
\bqa
\frac{d\sigma^{\mathrm{hard}}_{{ \chi_1}{ \chi_2}{}{}{} }}{ds'
        d\cos{\vartheta_3 }d\phi_3
        d\cos{\vartheta_5}}
=\alpha^3\frac{s-s'}{192\pi s^2}\frac{1}{\beta_e}
|\mathcal{H}^{\mathrm{hard}}_{{ \chi_1}{ \chi_2}{}{}{}} |^2,
\eqa
where $s'=(p_3+p_4)^2$  and 
\bqa
|\mathcal{H}^{\mathrm{hard}}_{ \chi_1 \chi_2{}{}{}} |^2 =
\sum_{\chi_3,\chi_4,\chi_5} |\mathcal{H}^{\mathrm{hard}}_{{ \chi_1}{ \chi_2}{ \chi_3}{ \chi_4}{\chi_5}} |^2.
\eqa

Here $\vartheta_5$ is the angle between $p_1$ and $p_5$ in the laboratory frame,
$\vartheta_3$ is the angle between 3-momenta $p_3$ and $p_5$ in the rest
frame of $(p_3 p_4)$-compound, $\phi_3$ is the azimuthal angle of the $p_3$ in the rest frame of
$(p_3 p_4)$-compound.

\subsection{The Born and virtual contributions}

The first step is the calculation of the Covariant Amplitude (CA)
and form factors (${\cal F}_{i}$).
In the system {\tt SANC} we calculate the CA of the annihilation to vacuum, i.e.
$2f2b \to 0$, and then turn over to the selected channel.
Then the helicity amplitudes (HAs) will be constructed.

\subsubsection{Covariant amplitude for the Born and virtual parts \label{Amplitudes}}
 
The covariant one-loop amplitude corresponds to the result of the
straightforward standard calculation  
of all diagrams contributing to a given process at the tree (Born) and 
one-loop  levels. 
The CA is represented in a certain basis made of strings of Dirac
matrices and/or  
external momenta (structures) contracted with polarization vectors of 
bosons $\varepsilon(p_i)$. The CA can be written in an explicit form using scalar form factors ${\cal F}_{i}$.
All masses, kinematic invariants and coupling constants, and other parameter
dependencies are included into these form factors, but tensor structures
with Lorenz indices made from strings of Dirac matrices are given by the~basis.
 
 Using the multichannel approach, we have found a complete massive basis and covariant amplitude.
The covariant amplitude for the processes we are interested in can be obtained
from Eq.~(\ref{uniCA}) by exploiting crossing symmetry.
We found 40 structures for the CA. 
By applying algebraic transformations, we  simplified the number of structures down to 24.
Using photon transversality, we obtained 6 ratios for vector and 5 for axial  form factors
${\cal F}_{i}$.
The final answer for the basis is 8 structures for the tensor and  4 structures for the pseudotensor.
In accordance with this,
the  next-to-leading order (NLO) EW RCs to the
process $2f 2\gamma \to 0$ can be parameterized in terms of 14 scalar form factors and the corresponding basic matrix elements,
8 vector and 4 axial ones.

For the covariant amplitude we have:
\bqa
  \label{uniCA}
  {\cal A}_{ l^+l^-\gamma\gamma} &=&
  \iap{p_1} 
  \Bigl[{\rm Str}^{v,0}_{\mu\nu} \left(\vmo {\cal F}^{0,v}(s,t,u) \right)
    + \sum_{j=1}^{7}{\rm Str}^{v,j}_{\mu\nu} {\cal F}^{j,v}(s.t.u)
    \nonumber\\
    && + \sum_{i=1}^{4}{\rm Str}^{a,i}_{\mu\nu} \gamma_5 {\cal F}^{i,a}(s,t,u)  
  \Bigr]
  \ip{p_2}\varepsilon^{\gamma}_{\nu}(p_3)\varepsilon^{\gamma}_{\mu}(p_4),
\eqa
with the structures 
\bqa
{\rm Str}^{v,0}_{\mu\nu} &=& 
\frac{s}{(\ml^2-t) (\ml^2-u)} \Bigl[
 -(i \sla{K_2}+\me) \tau_{\mu\nu}^6
 +2 i\frac{1}{\ml^2-u} \left(
  \tau_{\mu\nu}^7-\tau_{\mu\nu}^9
 -\frac{1}{2} (\sla{p_3}-\sla{p_4}) \tau_{\mu\nu}^{10} \right)\Bigr],
\nll 
{\rm Str}^{v,1}_{\mu\nu} &=& 
i  \sla{K_2}  \left(\frac{2}{s}\tau_{\mu\nu}^3+\tau_{\mu\nu}^{10}\right),
\nll
{\rm Str}^{v,2}_{\mu\nu} &=& 
- i \Bigl(
  \sla{K_2} (\kt^2 \tau_{\mu\nu}^3-\tau_{\mu\nu}^4)
 +\frac{1}{2} k_t\left[\tau_{\mu\nu}^1+2 i \me (\tau_{\mu\nu}^5+s\kt \tau_{\mu\nu}^{10})
            \right] \Bigr),
\nll                   
{\rm Str}^{v,3}_{\mu\nu} &=& 
i \left[
     \frac{1}{2}    \tau_{\mu\nu}^1
   + 2\sla{K_2} \kt \tau_{\mu\nu}^3
   + \left(-\sla{K_2}+i\me \right)\tau_{\mu\nu}^5
   + i\me s\kt                    \tau_{\mu\nu}^{10}\right],
\nll
{\rm Str}^{v,4}_{\mu\nu} &=& 
 \frac{4 i \me}{s}\sla{K_2} \tau_{\mu\nu}^{3} +\frac{s}{2}\tau_{\mu\nu}^{6}
+\sla{p_4} \tau_{\mu\nu}^7-\sla{p_3} \tau_{\mu\nu}^9-(\me^2+t) \tau_{\mu\nu}^{10},
\nll
{\rm Str}^{v,5}_{\mu\nu} &=& 
 -\tau_{\mu\nu}^0-\frac{2}{s}i \me \tau_{\mu\nu}^1-\frac{1}{2} (\me^2-t) \tau_{\mu\nu}^6
 -2\kt \left[ \frac{2}{s} i \me \sla{K_2}\tau_{\mu\nu}^3
        -\tau_{\mu\nu}^5-\frac{(3 \me^2-t)}{2} \tau_{\mu\nu}^{10}\right],
\nll
{\rm Str}^{v,6}_{\mu\nu} &=& 
\frac{s}{2}
\left(\frac{2}{s}\tau_{\mu\nu}^3 +\tau_{\mu\nu}^{10}\right),
\nll
{\rm Str}^{v,7}_{\mu\nu} &=& 
\tau_{\mu\nu}^4-\kt \left(\tau_{\mu\nu}^5
                      +\frac{\me^2-t}{2} \tau_{\mu\nu}^{10}\right),
\nll
    {\rm Str}^{a,1}_{\mu\nu} &=&
-i  \left( \sla{K_2} \tau_{\mu\nu}^6
 +\frac{4}{s}\kt \sla{K_2}\tau_{\mu\nu}^3
       -\left( k_t-i \frac{2\me}{s} \sla{p_4}\right) \tau_{\mu\nu}^7
       +\left( k_t+i \me \sla{p_3}\right) \tau_{\mu\nu}^9 \right),
\nll
    {\rm Str}^{a,2}_{\mu\nu} &=&
i  \sla{K_2} \left(\frac{2}{s}\tau_{\mu\nu}^3 + \tau_{\mu\nu}^{10}\right),
\nll
    {\rm Str}^{a,3}_{\mu\nu} &=&
-\frac{1}{2} \tau_{\mu\nu}^2 - i \sla{K_2}\left(\kt^2 \tau_{\mu\nu}^3-\tau_{\mu\nu}^4\right),
\nll
    {\rm Str}^{a,4}_{\mu\nu} &=&
-\frac{1}{2} \tau_{\mu\nu}^2 + i \sla{K_2} \kt \left(2 \kt \tau_{\mu\nu}^3+\tau_{\mu\nu}^5\right).
\eqa
where $\iap{p_1}$, $\ip{p_2}$ and $m_e$ are the bispinors and the mass of
the external fermions, respectively; 
$\varepsilon^{\gamma}_{\nu}(p_3)$ and $\varepsilon^{\gamma}_{\mu}(p_4)$ denote the photon polarization vector;
the vector and axial gauge-boson-to-fermion couplings are denoted by $v_l$ and $a_l$, respectively;
${\cal F}^{j,v}$ and ${\cal F}^{i,a}$ stand for the scalar form factors,
where ${\cal F}^{0,v}$ and ${\rm Str}^{0}_{\mu\nu}$ correspond to the lowest-order matrix element
and
\bqa
\sla{K_2} = \frac{1}{2}\left(\sla{p_3}-\sla{p_4}+\sla{p_2}-\sla{p_1}\right). \nonumber
\eqa

To obtain a compact form of the amplitude structures, we choose 10 auxiliary strings
\bqa
\tau_{\mu\nu}^0 &=& \gamma^\mu\sla{p_3}p_2^\nu+\sla{p_4}\gamma^\nu p_1^\mu,
\nll
\tau_{\mu\nu}^1 &=& s\Bigl[\gamma^\nu  (\ku p_1^\mu-\kt ( p_2^\mu - p_4^\mu))
                         -\gamma^\mu \left(\ku p_2^\nu-\kt(p_1^\nu-p_3^\nu)\right)\Bigr],
\nll
\tau_{\mu\nu}^2 &=& (\me \sla{K_2}-i t)
\left[  \gamma^\nu \left(p_1^\mu+k_t p_3^\mu\right)
       +\gamma^\mu \left(p_2^\nu+k_t p_4^\nu\right)\right],
\nll
\tau_{\mu\nu}^3  &=& p_3^\mu p_4^\nu,
\tau_{\mu\nu}^4   = p_1^\mu p_2^\nu,
\tau_{\mu\nu}^5   = p_1^\mu p_4^\nu+p_2^\nu p_3^\mu,
\nll
\tau_{\mu\nu}^6   &=& \gamma^\mu \gamma^\nu,
\tau_{\mu\nu}^7  = \gamma^\nu p_3^\mu,
\tau_{\mu\nu}^9  = \gamma^\mu p_4^\nu,
\tau_{\mu\nu}^{10}= \delta^{\mu\nu}.
\eqa
where $k_I=\frac{\ds \me^2-I}{\ds s}$, with $I=t,u$ (usual Mandelstam invariants).

In Eq.~(\ref{uniCA}), we keep the fermion mass in order to maintain photon
transversality without mass approximation.
Moreover, in the mass-containing denominators of ${\rm Str}^{0}_{\mu\nu}$, the mass
cannot be neglected because these denominators
correspond to the propagators of fermions that emit external photons
and thus lead to mass singularities.

The basic matrix elements, ${\rm Str}^{j}_{\mu\nu}$, are chosen to be explicitly
transversal in the photonic 4-momentum. 
That is, for all of them the following relations hold:
\bqa
    {\rm Str}^{j}_{\mu\nu} (p_3)_{\nu} = 0 ~\mbox{and}~ {\rm Str}^{j}_{\mu\nu} (p_4)_{\mu} = 0.
\eqa
We have checked that the form factors ${\cal F}^{j,v}$ and ${\cal F}^{i,a}$ are free of gauge parameters
and ultraviolet singularities; all calculations are made in the $R_\xi$ gauge.
The analytical expressions of the form factors
are too cumbersome to be presented in this paper.

\subsubsection{Helicity amplitudes for the Born and virtual parts}

\def\lcp {c^+}
\def\lcm {c^-}
\def\mel {m_l}
\def\izlll{z_{13}}
\def\izllll{z_{14}}
\def\sthlll{\sin{\vartheta}}
\def\cthlll{\cos{\vartheta}}
\def\sqrtLee{\sqrt{\lambda_{ee}}}
Using C, P, Bose and the transition between the final and initial symmetries, we write down four sets of HAs.
The presence of the electron masses gives additional terms proportional to
the factor $m_{l}$, which 
can be considered significant in calculations
at low energy.

\bqa
\mbox{\bf 1)}~
{\cal H}_{++-+} &=& -{\cal H}_{---+}(A=1),\quad {\cal H}_{+++-}=-{\cal H}_{---+}(A=-1),\\
{\cal H}_{--+-}&=&{\cal H}_{---+}(A=-1),\nonumber \\
{\cal H}_{---+} &=& -\frac{\sqrt{s}}{8}\lcp\lcm\Biggl\{
        s\mel\cthlll {\cal F}_{v_3}  \nonumber\\
                  &+&\sqrtLee \left[
    \mel\left(\frac{8}{ \izlll \izllll}{\cal F}_{v_1}+{\cal F}_{v_3}-4 {\cal F}_{v_4}\right)-2 {\cal F}_{v_6}+{\cal F}_{v_8}
   -A\mel \left({\cal F}_{a_3}+{\cal F}_{a_4}\right)\right]
            \Biggr\}. \nonumber
                                             \eqa
                                         
\bqa
\mbox{\bf 2)}~
{\cal H}_{-+\pm\pm} &=& {\cal H}_{-+--}(A=1),\quad
{\cal H}_{+-\pm\pm}={\cal H}_{-+--}(A=-1),
\\
{\cal H}_{-+--} &=& 
 \frac{1}{2}\sthlll \Biggl\{
          s \Bigl({\cal F}_{v_2}-\frac{s}{8}\left(1+\cos^2{\vartheta_{13}}\right){\cal F}_{v_3} 
             + A \cthlll \Bigl[({\cal F}_{a_1}+\frac{s}{4}  \left({\cal F}_{a_3}+{\cal F}_{a_4}\right)\Bigr]\Bigr)\nonumber \\
     &-&\sqrtLee \Bigl[
         s \cthlll \left(\frac{1}{4} {\cal F}_{v_3}-\frac{1}{2} {\cal F}_{v_4}+\frac{\mel}{s} {\cal F}_{v_6}\right) \nonumber \\
&-& A \Bigl({\cal F}_{a_1}-{\cal F}_{a_2}+\frac{s}{8}\left[(1+\cos^2{\vartheta_{13}}){\cal F}_{a_3}+2\cos^2{\vartheta_{13}}{\cal F}_{a_4} \right]\Bigr)
              \Bigr]
              \Biggr\}.\nonumber
              \eqa
    \bqa
\mbox{\bf 3)}~   
{\cal{H}}_{+--+} &=& {\cal{H}}_{-++-}(A=-1), \\
{\cal{H}_{-+-+}}&=&{\cal{H}}_{-++-}(A=1, \lcm \to -\lcp),
{\cal{H}_{+-+-}}={\cal{H}}_{-++-}(A=-1, \lcm \to -\lcp),
  \nonumber \\
{\cal H}_{-++-} &=&                 
     \frac{1}{8}\sthlll s \lcm\Biggl\{
         s \left[ \frac{1}{2}\lcm {\cal F}_{v3}+A( \frac{4}{s} {\cal F}_{a_1}+{\cal F}_{a_3}+{\cal F}_{a_4})\right] \nonumber \\
       &-& \sqrtLee \left(
       \frac{4}{ \izlll \izllll} {\cal F}_{v1}+{\cal F}_{v3}-2 {\cal F}_{v4}+\frac{4 \mel}{s} {\cal F}_{v6}
       + A (\frac{1}{2}\lcm {\cal F}_{a_3}-\cthlll {\cal F}_{a_4})
       \right)\Biggr\}.\nonumber \\
\mbox{\bf 4)}~
{\cal H}_{----} &=& -{\cal H}_{++++}(V=1),
\qquad
{\cal H}_{--++} = - {\cal H}_{++++}(\lcm \leftrightarrow \lcp, V=-1), \\
{\cal H}_{++--} &=& {\cal H}_{++++}(\lcm \leftrightarrow \lcp, V=-1), 
\nonumber \\
{\cal H}_{++++}  &=& -\frac{\sqrt{s}}{8}\Bigl\{  
       s\Bigl[ V\Bigl(\frac{8\mel}{\izlll\izllll} {\cal F}_{v_1}-2\lcp{\cal F}_{v_6}\Bigr) \nonumber \\
      &+& \cthlll \Bigl( \mel\left[\frac{8}{s} {\cal F}_{v_2} 
      - (4-\lcp\lcm){\cal F}_{v_3}+4{\cal F}_{v_4}\right]
        -4{\cal F}_{v_5}+2{\cal F}_{v_8}\Bigr)\Bigr]\nonumber \\
   &+&\sqrtLee \Bigl[
    \mel\left(\frac{8}{\izlll\izllll}{\cal F}_{v_1}+\frac{8}{s} {\cal F}_{v_2}
      -(4-3\lcp\lcm){\cal F}_{v_3}+4\cos^2{\vartheta_{13}}{\cal F}_{v_4}\right) \nonumber \\    
    &-& 4(1+\lcm) {\cal F}_{v_5}-2\lcp {\cal F}_{v_6}-4{\cal F}_{v_7}+2{\cal F}_{v_8}\Bigr]
   \Bigr\}\nonumber,
   \eqa
where $c^{\pm}=1\pm\cos\vartheta$.

\subsection{Real photon emission corrections} 
The real corrections consist of soft and hard radiative  contributions.
They are calculated using the bremsstrahlung modules. 
The soft bremsstrahlung has Born-like kinematics, 
while the phase space of hard radiation has an extra particle, photon.  

\subsubsection{Soft photon bremsstrahlung\label{HA_soft}}

The soft photon contribution contains
infrared divergences and has to compensate the corresponding divergences of one-loop virtual QED corrections.
It is factorized in front of the Born cross section. 
It depends on the auxiliary parameter
which separates the kinematic domains of the soft and hard photon emission in a given reference frame.
The polarization dependence is contained in $\sigma^{\rm Born}$. 

The explicit form is
\bqa
\sigma^{\rm soft} &=&
       - N \sigma^{\rm Born} 
       \frac{2}{\beta_e}
       \Biggl\{ 
2\left[\beta_e-
k
\ln x^2  \right]
\ln{\frac{2{\bar\omega}}{\lambda}} 
-\ln x +
k
     \left[\ln^2 x
     +\Litwo\left(1-1/x^2\right)
     \right] 
     \Biggr\}, \nonumber
\eqa
where
\bqa
N=\frac{\alpha}{2\pi} Q_e^2,\quad r_s=\frac{\ml}{s},\quad k=1-2 r_s, \quad 
x = \frac{1}{\sqrt{r_s}}\frac{1+\beta_e}{2}.
\nonumber
\eqa

Here ${\bar\omega}$ is the soft-hard separator, $\lambda$ is an auxiliary
infinitesimal photon mass, and
$\displaystyle{{L}_s = \ln{\frac {s}{m_e^2}}-1}$.

\subsubsection{Hard photon bremsstrahlung\label{HA_hard}}

Spin effects of hard photon bremsstrahlung for photon-pair production using the 
method of helicity amplitudes were investigated  in
\cite{Dittmaier:1998nn}, \cite{Shishkina:2002ww}.
In the presented results, we used our universal
massive module for hard photon bremsstrahlung for
$l^+l^-\gamma\gamma\gamma \to 0$ by appropriately unfolding it in
channel (\ref{cppp}),
where $0$ stands for {\em vacuum}, and  all masses are not neglected.

The field strength bivector is an antisymmetric tensor and can naturally be expressed as an element of the Clifford algebra of Dirac matrices by contracting with $\gamma^{[\mu}\gamma^{\nu]}=\gamma^{\mu}\wedge\gamma^{\nu}$.

Let us consider a photon with 4-momentum $k^2=0$ and polarization vector $\varepsilon$. The Maxwell bivector (contracted with Dirac matrices) is
\begin{align*}
\mathbf F &\equiv F_{\mu\nu}\gamma^{\mu}\gamma^{\nu} 
= \slashed k \wedge \slashed \varepsilon.
 \end{align*}
The Maxwell equation becomes $\slashed k \mathbf{F} = 0$. It is also evident that  gauge transfomation $\varepsilon\to\varepsilon + C k$  leaves the bivector $\mathbf{F}$ unaffected.

The axial gauge can be defined by the additional condition $\varepsilon \cdot g = 0$ with some (massive) vector $g$. Solving it together with $\varepsilon\cdot k =0$, we obtain a polarization vector in the axial gauge 
\begin{align*}
\slashed \varepsilon &=\dfrac{\grade{1}{\slashed  g \mathbf F} }{g\cdot k} 
,&
\grade{1}{A} &\equiv \Trbar[A\gamma^{\mu}]\gamma_{\mu}
,&
\slashed \varepsilon(g_1)-\slashed \varepsilon(g_2) &=
-\dfrac{\Trbar[\slashed g_1 \slashed g_2 \mathbf{F} ] }{( g_1\cdot k)( g_2\cdot  k)} \slashed k
,&
\Trbar = \dfrac{1}{4}\Tr
.
\end{align*} 
Changing the vector $g$ leads to gauge transformation.

The helicity amplitude for hard photon bremsstrahlung
{ is organized as a sum of three cyclically symmetric terms}
\begin{align*}
\mathcal{A} &= 2\sqrt{2}(\mathcal{A}^3 + \mathcal{A}^4 + \mathcal{A}^5)
,&
\mathcal{A}_3 = \mathcal{A}_5\big|_{5\to 3\to 4\to 5}
,&
\mathcal{A}_4 = \mathcal{A}_5\big|_{5\to 4\to 3\to 5}
.
\end{align*}

So it is enough to consider only the single term. 
The Maxwell bivector for helicity states can be factorized $ \mathbf{F}_5^{\chi_5} = u_5^{\chi_5}\bar{v}_5^{\chi_5}$, and the corresponding term decays into building blocks:
\begin{align*}
  \mathcal{A}^5_{\xi_1\xi_2\chi_3\chi_4\chi_5}  &= 
  R_{\xi_1}{}^{\chi_1}(1)R_{\xi_2}{}^{\chi_2}(2)
  \dfrac{
  -\mathcal{S}^{5}_{\chi_5} \mathcal{B}_{\chi_1\chi_2\chi_3\chi_4} 
  + \mathcal{C}^5_{\chi_1\chi_5} \mathcal{G}^{5}_{\chi_5\chi_2\chi_3\chi_4}}{z_{23}z_{24}} ,
\end{align*} 
\begin{align*}
&\mathcal{S}^{5}_{\chi_5} = \dfrac{\Tr[\slashed{p}_1 \slashed{p}_2\mathbf{F}_5] }{\sqrt{2}z_{15} z_{25}}
 ,&
 \mathcal{B}^{5}_{\chi_1\chi_2\chi_3\chi_4} &=  
 \bar{v}_1  \Big(  \grade{1}{\mathbf{F}_3 \slashed{p_2}   \mathbf{F}_4}
		- \slashed{p_2} \grade{0,4}{\mathbf{F}_3   \mathbf{F}_4} 
		\Big) u_2  
,\\
&\mathcal{C}^5_{\chi_1\chi_5} =  
 \dfrac{\bar{v}_1 u_5}{z_{15}}
,&
\mathcal{G}^{5}_{\chi_5\chi_2\chi_3\chi_4} &=  
 \bar{v}_5  \Big(
	  \grade{1}{\mathbf{F}_3 \slashed{p_2}   \mathbf{F}_4}
	-  \slashed{p_2} \grade{0,4}{\mathbf{F}_3   \mathbf{F}_4} 
		\Big) u_2 
,
\end{align*} 
 with the abbreviations $u_i \equiv u^{\chi_i}(p_i)$, $\bar v_i \equiv \bar v^{\chi_i}(p_i)$, $\mathbf{F}_j\equiv \mathbf{F}^{\chi_j}(p_j)$, and $\grade{0,4}{A} = \Trbar[A]+\Trbar[A\gamma_{5}]\gamma_{5}$.

We work in the chiral representation of gamma-matrices and exploit Weyl spinors. 
To decompose the Dirac spinors into Weyl components, we use the following notation:
\begin{gather*}
\begin{aligned}
\slashed{p} &= 
\begin{pmatrix}
   & \check{p} \\
\hat{p} &  
\end{pmatrix} 
,&
u  &=
\begin{pmatrix}
 \SpIA{u} \\   \SpIB{u} 
\end{pmatrix}
,&
\mathbf{F} &=
\begin{pmatrix}
\check{\mathbf{ F}} & \\ & \hat{\mathbf{F}}
\end{pmatrix}
,&
\bar{v}  &= 
\Big(\begin{matrix}
 \SpAI{\bar{v}} ,&   \SpBI{\bar{v}} 
\end{matrix}\Big),
\end{aligned}
\end{gather*} 
For the massless particle with momentum $p_i$, we have 
$ \check{p}_i = \SpIA{i}\SpBI{i}$,  		$\hat{p}_i = \SpIB{i}\SpAI{i}$. For the massive particle with $p_i^2=m_i^2$, we use the projection on the light-cone of some auxiliary momentum. To evaluate the term $\mathcal{A}^5$ of the amplitude, we find   that one of the most 
economical choices is  to use $p_5$:
\begin{align*}
\hat{k}_i &= \dfrac{\hat{p}_i \check{p}_5 \hat{p}_i}{2p_i\cdot p_5} 
,&
\hat{k}_i &=\SpIB{i}\SpAI{i}
,&
\SpIA{i} &= \dfrac{\check{p}_i\SpIB{5}}{\SpBIB{i}{5}}
,&
\SpIB{i} &= \dfrac{\hat{p}_i\SpIA{5}}{\SpAIA{i}{5}}
\end{align*}
The Dirac solutions in terms of spinors for $k_i$  are
	\begin{align*}
	&\begin{aligned}  
 		u_{i}^{+} &=  \begin{bmatrix}\SpIA{i} \\ \dfrac{m_i}{\SpBIB{i}{5}}\SpIB{5}\end{bmatrix}
		,&
 		u_{i}^{-}  &= \begin{bmatrix} \dfrac{m_i}{\SpAIA{i}{5}}\SpIA{5}\\\SpIB{i} 
 		\end{bmatrix}
	\end{aligned}	
,&
	&\begin{aligned}  
 		\bar{v}_{i}^{+}  &= \begin{bmatrix} 
 		\SpAI{i} ,&- \dfrac{m_i}{\SpBIB{i}{5}}\SpBI{5}
 		\end{bmatrix}
		,&
 		\bar{v}_{i}^{-}  &= \begin{bmatrix} 
 		 - \dfrac{m_i}{\SpAIA{i}{5}}\SpAI{5} ,& \SpBI{i} 
 		\end{bmatrix}
	\end{aligned} 
	\end{align*}  

The explicit expressions of the amplitude components $\mathcal{A}$ 
are written as follows:
\begin{align*}
&\mathcal{B}^{5}_{\chi_1\chi_2\chi_3\chi_4} =  
\begin{bmatrix}
  m\SpBIB{\bar{v}_1}{ u_2} \SpAIA{3}{4}^2 & \SpAIIB{3}{p_2}{4}\big(\SpAIA{\bar{v}_1}{3}\SpBIB{4}{u_2}+\SpBIB{\bar{v}_1}{4}\SpAIA{3}{u_2}\big) \\
 \SpBIIA{3}{p_2}{4}\big(\SpBIB{\bar{v}_1}{3}\SpAIA{4}{u_2}+\SpAIA{\bar{v}_1}{4}\SpBIB{3}{u_2}\big) & 
 m\SpAIA{\bar{v}_1}{ u_2} \SpBIB{4}{3}^2
\end{bmatrix}_{\chi_3\chi_4},
\end{align*}
\begin{align*}
&\mathcal{G}^{5}_{\chi_5\chi_2\chi_3\chi_4} =  
\begin{bmatrix}
  m\SpBIB{\bar{v}_5}{ u_2} \SpAIA{3}{4}^2 & \SpAIIB{3}{p_2}{4}\big(\SpAIA{\bar{v}_5}{3}\SpBIB{4}{u_2}+\SpBIB{\bar{v}_5}{4}\SpAIA{3}{u_2}\big) \\
 \SpBIIA{3}{p_2}{4}\big(\SpBIB{\bar{v}_5}{3}\SpAIA{4}{u_2}+\SpAIA{\bar{v}_5}{4}\SpBIB{3}{u_2}\big) &
  m\SpAIA{\bar{v}_5}{ u_2} \SpBIB{4}{3}^2
\end{bmatrix}_{\chi_3\chi_4},
\end{align*}

\begin{align*} 
\mathcal{S}^{5}_{\chi_5} &=   \begin{bmatrix} -\dfrac{ \SpBIB{1}{2} }{ \SpBIB{1}{5}  \SpBIB{2}{5}}
&
-\dfrac{ \SpAIA{1}{2} }{ \SpAIA{1}{5}  \SpAIA{2}{5}}
\end{bmatrix}_{\chi_5}
,&   
\mathcal{C}^5_{\chi_1,\chi_5} &=    \dfrac{1}{z_{15}}\begin{bmatrix}
  \SpAIA{1}{5} &   0  \\
  0 &   \SpBIB{1}{5}  
 \end{bmatrix}_{\chi_1\chi_5},
\end{align*}
\begin{gather*} 
\begin{aligned}
 \SpAIA{\bar{v}_i}{u_j} &=  
 \begin{bmatrix}
  \SpAIA{i}{j}                         & m_j\dfrac{\SpAIA{i}{5}}{\SpAIA{j}{5}} \\
m_i\dfrac{\SpAIA{5}{j}}{\SpAIA{5}{i}} & 0
\end{bmatrix}_{\chi_i\chi_j}
,&
 \SpBIB{\bar{v}_i}{u_j} &= 
 \begin{bmatrix}
           0              & m_i\dfrac{\SpBIB{j}{5}}{\SpBIB{i}{5}} \\
m_j\dfrac{\SpBIB{5}{i}}{\SpBIB{5}{j}} &  \SpBIB{i}{j}  
\end{bmatrix}_{\chi_i\chi_j}.
\end{aligned}
\end{gather*}

From the momentum $ p_i = \{E_i,p_i^x,p_i^y,p_i^z\}$ with $p_i^2 =m_i^2$ there can be built two massless vectors $  k_{i^*} = \{|\vec{p}_i|,-p_i^x,-p_i^y,-p_i^z\}$ and $k_{i^\flat} = p_i-\frac{\ds m_i^2}{\ds 2p_i\cdot k_{i^*}}k_{i^*}$ , with $k_{i^*}^2 =k_{i^\flat}^2 =0$. The corresponding spinors $\SpIA{i^{*}}$ and $\SpIA{i^\flat}$ allow evaluating the rotation matrix
\bqa
R_{\xi_i}^{\phantom{a}\chi_i}(i)  
&=& \left[\begin{matrix}\displaystyle
\frac{\SpBIB{i^{\flat}\!}{5}}{\SpBIB{i}{5}}
&\displaystyle
\frac{m_i\SpAIA{i^*}{5}}{\SpAIA{i^*}{i^{\flat}}\SpAIA{i}{5}}
\\\displaystyle
\frac{m_i\SpBIB{i^*}{5}}{\SpBIB{i^*}{i^{\flat}}\SpBIB{i}{5}}
&\displaystyle
\frac{\SpAIA{i^{\flat}}{5}}{\SpAIA{i}{5}}
\end{matrix}\right]
.
\nonumber
\eqa

\section{Numerical results \label{Sect_Num}}

In this section, we firstly present comparison of the result obtained by means of the {\tt SANC} system with the
tree-level results for Born and hard photon bremsstrahlung of the \CalcHEP{} \cite{Belyaev:2012qa} and \WHIZARD{} 
\cite{Ohl:2006ae, Kilian:2007gr,Kilian:2018onl} codes.
Also, the NLO QED RCs are compared with the \BabaYaga code~\cite{CarloniCalame:2019nra} 
and weak corrections with those published 
in~\cite{CarloniCalame:2019dom}.

In the second part of the section, we show the predictions for NLO EW RCs obtained by the {\tt SANC} program.  

If not specified separately, the following
set of input parameters is used:
\begin{tabbing}
\hspace{3em}\=\hspace{2em}\=\hspace{10em}\=\hspace{3em}\=\hspace{2em}\=\hspace{1em}\kill
$\alpha^{-1}(0)$\> = \> 137.035999084 \> \\
$M_W$ \>=\> 80.379 GeV\> $\Gamma_W$ \>=\> 2.0836 GeV\\
$M_Z$ \>=\> 91.1876 GeV\> $\Gamma_Z$  \>=\> 2.4952 GeV \\
$M_H$ \>=\> 125.0 GeV\> $m_e$ \>=\> 0.51099895000 MeV \\
$m_\mu$  \>=\> 0.1056583745 GeV\>  $m_\tau$  \>=\> 1.77686 GeV \\
$m_d$ \>=\> 0.083 GeV\> $m_s$  \>=\> 0.215 GeV \\
$m_b$  \>=\> 4.7 GeV\>  $m_u$  \>=\> 0.062 GeV \\
$m_c$  \>=\> 1.5 GeV\>  $m_t$  \>=\> 172.76 GeV 
\label{parameters}
\end{tabbing}
\text{with the angular cuts for at least two photons:}
$|\cos{\theta_\gamma}| < 0.9$.
In practical calculations we used $\Gamma_W = \Gamma_Z = 0$.

\subsection{Comparison with other codes}

Firstly, we have compared the results for the  
Born cross section for several c.m. energies ($\sqrt{s}=250,500,1000$ GeV)
and the degree of initial beam polarization. The agreement in 
5 digits was found, so we omitted the corresponding table.

Secondly, we have compared the results for  the
hard photon bremsstrahlung cross section for the same c.m. energies
with the {\CalcHEP} and {\WHIZARD} codes.
The results are given within the $\alpha(0)$ EW scheme in 
Table~\ref{Table:tuned_hard_SCW}.
For the cross sections, an additional cut on the
photon energy $E_\gamma \ge \omega = 10^{-4} \sqrt{s}/2$ is applied.
At least two photons lie in $|\cos{\theta_\gamma}| < 0.9$.
The comparison demonstrates very good (within four to five digits) agreement with the above-mentioned codes

\begin{table}[ht!]
\centering
\caption{
The triple tuned comparison between the {\SANC} (first line), {\CalcHEP} (second line) and {\WHIZARD} (third line)
results for the hard bremsstrahlung contributions to polarized $e^+e^- \to \gamma \gamma(\gamma)$ process}
\label{Table:tuned_hard_SCW}
    \begin{tabular}{llll}
    \hline\hline
    $\sqrt{s}$, GeV & {250} & {500} & {1000} \\
    \hline
    {\tt SANC}	              & 4.467(2)      & 1.177(1)    & 0.3095(1)  \\  
    {\tt CalcHEP}             & 4.465(1)   & 1.177(1) & 0.3096(1) \\        
	{\tt WHIZARD}             & 4.465(1)   & 1.180(1) & 0.3097(1) \\
	\hline 
\end{tabular} 
\end{table}

We also compared the NLO QED calculations between 
the {\tt SANC} and \BabaYaga codes.
In Tables ~\ref{tab:sancvsbabayaga1} and \ref{tab:sancvsbabayaga2}, 
we present a tuned comparison 
of the integrated cross sections produced for
for two c.m. energy regions: 
low ($\sqrt{s} = 1$ and 10~GeV) and 
high  ($\sqrt{s} = 91, 160, 240$ and $365$ GeV) energies
with original setups and cuts~(for details, see~\cite{CarloniCalame:2019nra,CarloniCalame:2019dom}).

\begin{table}[h!]
    \centering
    \caption{
    The tuned comparison of the Born and NLO QED
    integrated cross sections produced by the ${\tt SANC}$ and ${\tt BabaYaga}$ codes
    at low energies
\label{tab:sancvsbabayaga1}}    
    \begin{tabular}{cll}
    \hline\hline
    $\sqrt{s}$, GeV	    & 1		&    10 \\
    \hline
    \multicolumn{3}{c}{Born, nb}\\
    \SANC               & 137.532(1) & 1.3755(1)\\
    \BabaYaga           & 137.53     & 1.3753\\
    \multicolumn{3}{c}{NLO QED, nb}\\
    \SANC               & 129.46(2)  & 1.2623(3)\\
    \BabaYaga           & 129.45     & 1.2620\\
    \hline\hline
    \end{tabular}
\end{table}

\begin{table}[ht]
\caption{
Tuned comparison  of the Born and NLO QED
integrated cross sections produced by the ${\tt SANC}$ and ${\tt BabaYaga}$ codes
at high energies
 \label{tab:sancvsbabayaga2}}
    \centering
    \begin{tabular}{cllll}
    \hline\hline
    $\sqrt{s}$, GeV	 & 91		&    160     & 240       & 365 \\
    \hline
    \multicolumn{5}{c}{Born, pb}\\
    \SANC            & 39.822(1) & 12.884(1) & 5.7252(1) &  2.4758(2) \\
    \BabaYaga        & 39.821    & 12.881    & 5.7250    &  2.4752\\
    \multicolumn{5}{c}{NLO QED, pb}\\
    \SANC            & 41.04(1)  & 13.289(3) & 5.907(1)  & 2.556(1)\\
    \BabaYaga        & 41.043    & 13.291    & 5.9120    & 2.5581 \\
    \hline\hline
    \end{tabular}
\end{table}

\begin{figure}[h!]
    \centering    
    \includegraphics[scale=0.9]{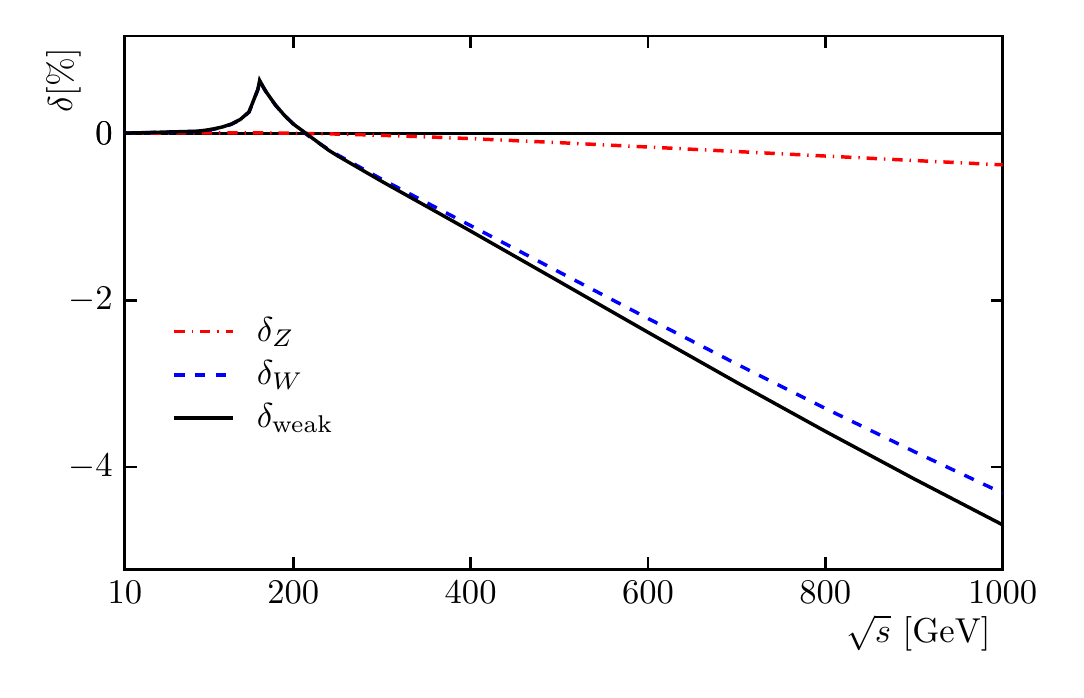}
    \includegraphics[scale=0.9]{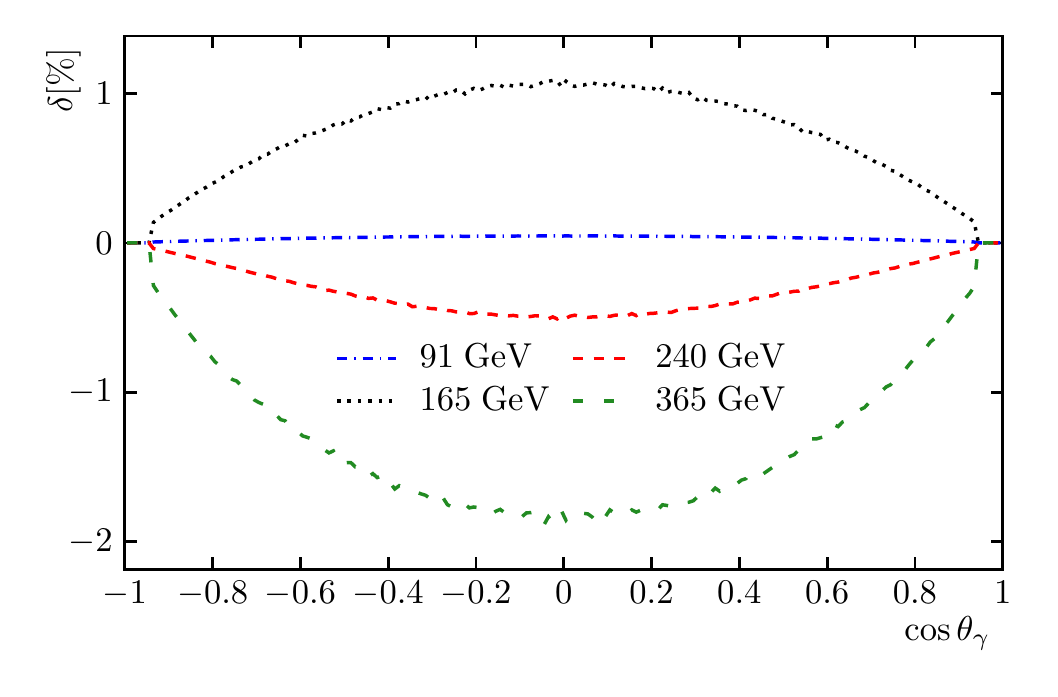}
    \caption{Upper plot: the integrated relative contributions of $Z$ and $W$ bosons
    to weak RCs.
    Lower plot: the differential relative weak RCs for
    c.m. energies at FCCee
    }
    \label{fig:delta-weak-int}
\end{figure}

Tables~\ref{tab:sancvsbabayaga1} and \ref{tab:sancvsbabayaga2} show perfect agreement of the NLO QED
results (within the statistical errors) and we consider these corrections are under control.

To compare the weak part of the NLO RCs, we have produced energy and angular distributions of the
relative corrections $$\delta = \sigma^{\text{1-loop}}/\sigma^{\text{Born}}-1, \%$$
which are presented in Fig.~\ref{fig:delta-weak-int}.
In the upper panel, the separate contributions for virtual $Z$ and $W$ boson contributions and
their sum are shown as a function of unpolarized beams. In the lower panel the
angular distributions for several c.m. energies are given.

The obtained RCs show very good qualitative agreement with those given in~Fig.~3 of~\cite{CarloniCalame:2019dom}.

\subsection{Born, one-loop cross sections and relative corrections}

In this part of the section,
we give our results for the Born, one-loop cross sections and relative corrections
\footnote{We investigate the energy range $250-1000$ GeV, because  the ILC collider was originally proposed to run at a cms energy 
$\sqrt{s} = 500$ GeV \cite{Bambade:2019fyw},
and recent scenarios with $\sqrt{s} = 250$ GeV and $\sqrt{s} = 1$ TeV \cite{Fujii:2017vwa,Zarnecki:2020ics} were also considered.}.
They were calculated with the parameters~(\ref{parameters}) and the following set
of the electron $(P_{e^-})$ and positron $(P_{e^+})$
beam polarization:
\bqa \label{SetPolarization}
&& (P_{e^-}, P_{e^{+}}) =
(0,0),\;(0.8,0.3),\;(-0.8,0.3).
\eqa

\subsubsection{Energy dependence}

In Tables~\ref{Table-pol-250} - \ref{Table-pol-1000},
the results of the integrated Born and one-loop cross sections in pb
and the relative corrections in percent 
are presented
separably for NLO QED and weak RC.
The results are given for the c.m. energies $\sqrt{s}=250, 500, 1000$ GeV 
and for degrees (\ref{SetPolarization}) of the initial particle polarization 
in the $\alpha(0)$ EW scheme.

As it is seen from the tables, the cross sections and the
weak RCs are sensitive to the degree of the initial beam polarization
while the QED RCs are rather flat.
For c.m. energy $\sqrt{s}=250$ GeV the weak RCs are negative and relatively 
small compared to QED RCs (approximately $5-6$ times). 
As for c.m. energy $\sqrt{s}=1000$ GeV the
weak RCs become compatible with the QED RCs in the unpolarized case ($6.9\%$ vs $-5.1\%$)
and even larger ($6.9\%$ vs $-8.6\%$) for polarization $(P_{e^-},P_{e^+})=(-0.8,0.3)$ 
but with the opposite sign. This means that the weak RCs dominates at high energies and 
must be taken into account.

\begin{table}[ht!]
\caption{
Born cross section $\sigma$ (pb), NLO QED and weak relative correction
$\delta$ (\%) for the c.m. energy $\sqrt{s}$ = 250 GeV
and the set (\ref{SetPolarization}) of the polarization degree 
of the initial particles
}
\label{Table-pol-250}
    \centering
      \begin{tabular}{lccc}
            \hline\hline
            $P_{e^-}, P_{e^+}$ & $ 0$, $ 0$ & $0.8$, $0.3$ & -$0.8$, $0.3$\\
            \hline
            $\sigma^{\text{Born}}$,~pb & $4.2617(1)$ & $3.2388(1)$ & $5.2845(1)$\\
            $\sigma^{\text{QED}}$,~pb
            & 4.535(2) & 3.4488(5) & 5.619(1)\\
            $\delta^{\text{QED}}$,\%
            & 6.42(4) & 6.48(1) & 6.32(2)\\
            $\sigma^{\text{weak}}$,~pb & $4.2481(1)$ & $3.2345(1)$ & $5.2544(1)$\\
            $\delta^{\text{weak}}$,\% & $-0.32(1)$ & $-0.13(1)$ & $-0.57(1)$\\
            \hline
            \hline
        \end{tabular}
\end{table}            
            
\begin{table}[ht!]
\caption{
The same as in Tab.~\ref{Table-pol-250}
but for the c.m. energy $\sqrt{s}$ = 500 GeV}
\label{Table-pol-500}
    \centering
      \begin{tabular}{lccc}
            \hline\hline
            $P_{e^-}, P_{e^+}$ & $ 0$, $ 0$ & $0.8$, $0.3$ & -$0.8$, $0.3$\\
            \hline            
            $\sigma^{\text{Born}}$,~pb & $1.06542(1)$ & $0.80972(1)$ & $1.32112(1)$\\
            $\sigma^{\text{QED}}$,~pb 
            & 1.1365(2) & 0.8641(1) & 1.4085(3)\\
            $\delta^{\text{QED}}$,\% 
            & 6.67(2) & 6.72(2) & 6.62(2)\\
            $\sigma^{\text{weak}}$,~pb & $1.04396(1)$ & $0.81165(1)$ & $1.25437(1)$\\
            $\delta^{\text{weak}}$,\% & $-2.01(1)$ & $0.24(1)$ & $-5.05(1)$\\       
            \hline
            \hline
        \end{tabular}
\end{table}            
            
\begin{table}[ht!]
\caption{
The same as in Tab.~\ref{Table-pol-250}
but for the c.m. energy $\sqrt{s}$ = 1000 GeV}
\label{Table-pol-1000}
    \centering
      \begin{tabular}{lccc}
            \hline\hline
            $P_{e^-}, P_{e^+}$ & $ 0$, $ 0$ & $0.8$, $0.3$ & -$0.8$, $0.3$\\
            \hline            
            $\sigma^{\text{Born}}$,~pb & $0.266353(1)$ & $0.202429(1)$ & $0.330279(1)$\\
            $\sigma^{\text{QED}}$,~pb 
            & 0.28474(5) & 0.21661(4) & 0.3531(1)\\
            $\delta^{\text{QED}}$,\% 
             & 6.90(2) & 7.00 (2) & 6.90(4)\\
            $\sigma^{\text{weak}}$,~pb & $0.252650(1)$ & $0.197583(1)$ & $0.301040(1)$\\
            $\delta^{\text{weak}}$,\% & $-5.14(1)$ & $-2.39(1)$ & $-8.85(1)$\\       
            \hline
            \hline 
        \end{tabular}
  \end{table}
  
\begin{figure}[!h]
\begin{center}
\includegraphics[scale=0.9]{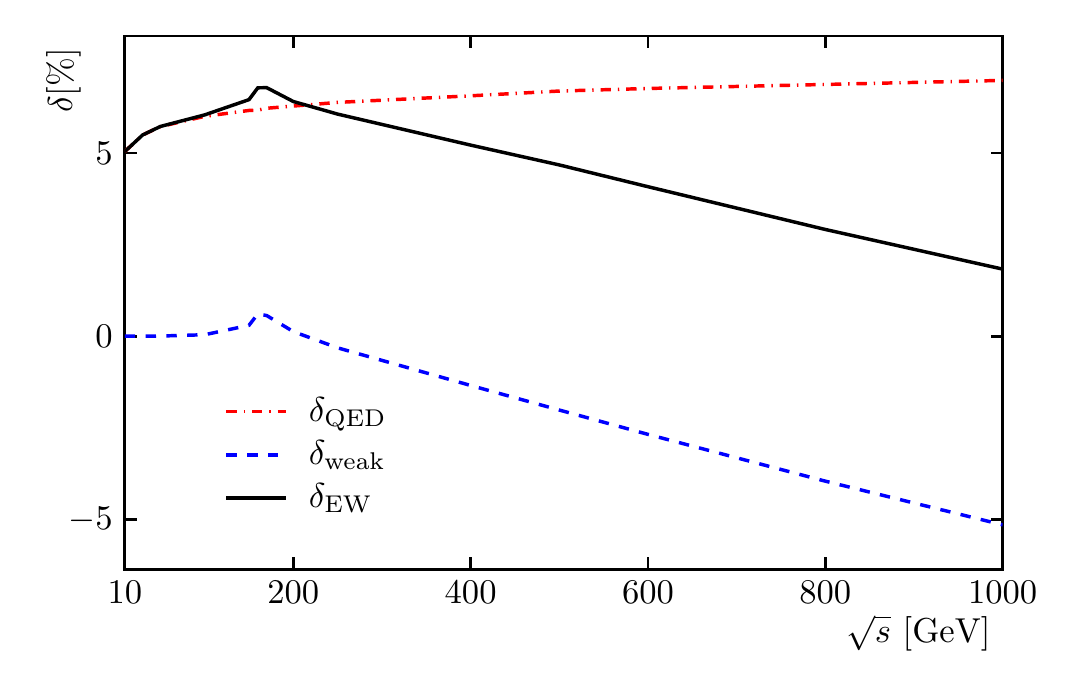}
\caption{
The unpolarized NLO QED, weak and NLO EW relative correction
$\delta$ (\%) for the c.m. energy range $\sqrt{s}$ = $10-1000$ GeV
\label{delta-nlo-sq}}
\end{center}
\end{figure}

To demonstrate the interference of the QED and weak RCs, we plotted the energy scan.
Figure ~\ref{delta-nlo-sq} shows the unpolarized QED, weak and summary (EW = QED + weak)
relative correction $\delta$ (\%) for the c.m. energy range $\sqrt{s} = 10-1000$ GeV.
In the calculations, only angular cuts for at least two photons were applied.
As one can see from the picture, the QED RCs dominate in the energy range
up to $\sqrt{s} = 100$ GeV. In the range $\sqrt{s} = 100-200$ GeV, the weak contribution is
positive and increase the NLO RCs. Then the weak relative corrections become negative
and start to reduce the total RCs to approximately 2\% at $\sqrt{s} = 1000$ GeV. 
One can see that, the weak corrections change $\delta_{\rm EW}$ drastically
for high energies (starting approximately from $\sqrt{s} = 200$ GeV).
It should also be stressed that additional kinematical cuts such as, for example,
photon energy cut, reduce the magnitude of the QED RCs and dominance of the weak RCs 
becomes harder (stronger?).

\subsubsection{Left-right asymmetry}

\begin{figure}[!h]
\begin{center}
\includegraphics[scale=0.9]{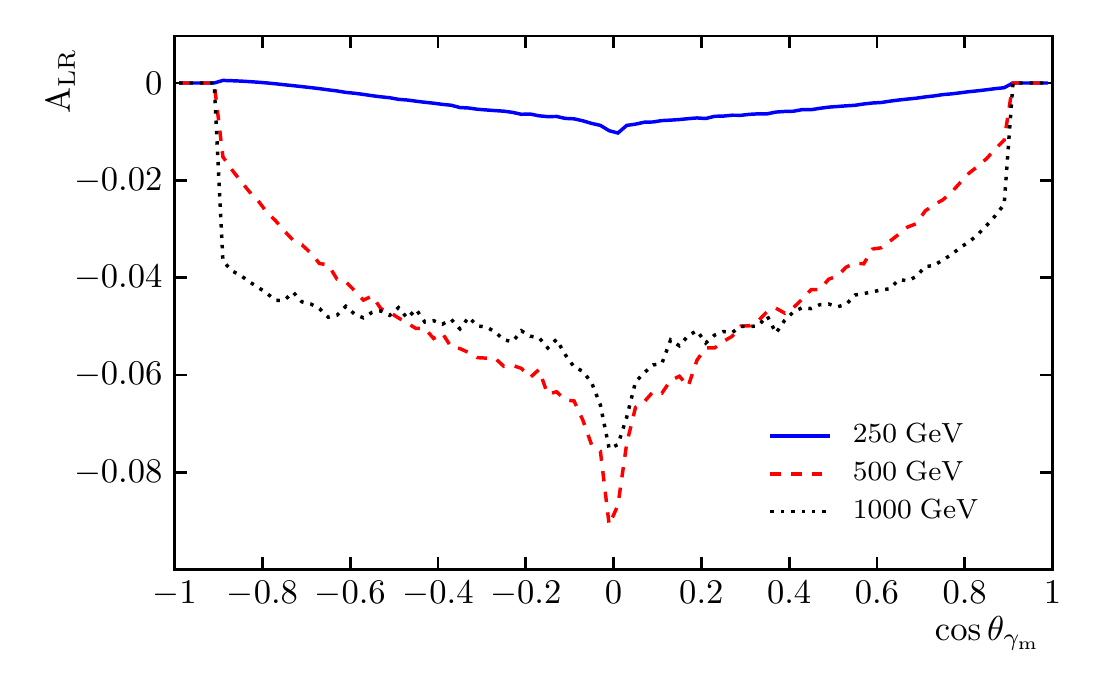}
\caption{
Angular distributions for $A_{\rm LR}$ asymmetry
over the highest energy photon angle
at several c.m. energies
\label{alr}
}
\end{center}
\end{figure}

The left-right asymmetry is also calculated  at c.m. energies 
$\sqrt{s}=250,500,1000$ GeV,
and the angular distributions are shown in Figure~\ref{alr}.

The $A_{LR}$ is defined in the following form:
\bqa
A_{LR}=\frac{\sigma_{LR}-\sigma_{RL}}{\sigma_{LR}+\sigma_{RL}},
\eqa
where $\sigma_{LR}$  and $\sigma_{RL}$ are the cross sections for 100$\%$
polarized electron-positron $e^-_{L}e^+_{R}$ and  $e^-_{R}e^+_{L}$ 
initial states.

As it is seen from the figure, the angular dependence of the
asymmetry is very weak at $\sqrt{s}=250$ GeV but become stronger at
$\sqrt{s}=1000$ GeV. 
This reaction does not have any clearly seen resonance
(in contrast with the $s$-channel pair-lepton production, for example, 
where the $Z$ boson defines the peak of the cross section). The asymmetry
does not give any experimental information on the mixing angle
$\sin^2\theta_W$
and just shows an order of parity violation.

\section{Conclusions and outlook \label{Sect_Concl}} 
 
In this paper, we have considered the complete one-loop 
electroweak corrections to the process
of the polarized electron-positron annihilation into a photon pair
within the {\tt SANC} system framework.
The helicity amplitudes were used for the Born and virtual parts
as well as for the real photon emission (soft/hard bremssrahlung)
taking into account the masses of the initial particles.

The numerical results were evaluated 
within the {\tt SANC} system framework in the $\alpha(0)$ scheme for c.m. energies from 10~GeV  
to about 1000~GeV which are relevant for the existing and future $e^+e^-$ colliders.
We reached excellent agreement at the tree level for the Born and hard 
photon bremssrahlung  
between {\tt SANC},
{\tt CalcHEP} 
and
{\tt WHIZARD}. 

At the one-loop level for unpolarized beams we have compared the obtained
results with external codes.
Firstly, we performed a tuned comparison of the NLO QED corrections
with the \BabaYaga{} code and found very good numerical agreement.
Secondly, we obtained good qualitative agreement of the weak radiative corrections 
with the figures given in the world literature.

We have presented the electroweak radiative corrections
impacting the Born and complete one-loop cross sections as well as relative corrections at c.m energies $\sqrt{s} = 250, 500, 1000$ GeV.
The results are given for unpolarized and polarized cases and
demonstrate the strong dependence of the total/differential cross section and relative corrections on the polarization effects.

We would like to emphasize that weak effects give large negative 
corrections and totally compensate the QED radiative corrections at high energies and therefore must be taken into account.

Analytical calculations for all parts of the cross section
were performed for the case of annihilation into vacuum
in the massive case $2f 2\gamma \to 0$. 
This lays the foundation 
for calculating all cross channels.

Considering the $e^+ e^- \to \gamma \gamma $
process as one for luminometry propose, 
one needs to take into account
high-order effects, such as leading multi-photon QED
logarithms and leading two-loop corrections.
This is the forthcoming part of our work on this process.

\section{Funding}
\label{sec:funding}
The research is supported by the Russian Science Foundation (project No. 22-12-00021).


\providecommand{\href}[2]{#2}
\end{document}